\documentstyle[epsf,tighten,aps,pra]{revtex}

\begin{document}

\protect\twocolumn[

\title{Quantum identification system}

\author{Miloslav Du\v sek$^1$, Ond\v rej Haderka$^{2,1}$, Martin
Hendrych$^{2,1}$, Robert My\v ska$^{2,1}$\vspace{1cm}}
\address{$^1$Department of Optics, Palack\' y University, 17.~listopadu 50,
         772~00 Olomouc, Czech Republic}
\address{$^2$Joint Laboratory of Optics of Palack\' y University and the
Physical Institute of the Czech Academy of Sciences,\\
17.~listopadu 50, 772~00 Olomouc, Czech Republic}

\begin{center}\begin{minipage}{0.9\hsize}\small
A secure quantum identification system combining a classical
identification procedure and quantum key distribution is proposed. Each
identification sequence is always used just once and new sequences
are ``refuelled'' from a shared provably secret key transferred through
the quantum channel. Two identification protocols are devised. The first
protocol can be applied when legitimate users have an unjammable public
channel at their disposal. The deception probability is derived for the case
of a noisy quantum channel. The second protocol employs
unconditionally secure authentication of information sent over the public
channel, and thus it can be applied even in the case when an adversary is
allowed to modify public communications. An experimental
realization of a quantum identification system is described.\\
{PACS 03.67D}\\
\end{minipage}\end{center}
\maketitle
]


\section{Introduction}

Electronic communications have become one of the main pillars of the
modern society. Their utilization places new demands on the
establishment of security of transmitted data. In everyday life there
are many situations when it is necessary to conceal the contents of
information conveyed over insecure communications lines, such as when
databases containing confidential data on citizens are to be distributed
among authorities, when financial transactions are performed between
banks (or for electronic shopping over the Internet), or when we want to
withdraw money from automated teller machines, and, of course, for
military and diplomatic purposes.

In all these instances, cryptography proves very helpful. One of the
basic cryptographic tasks is to certify the identities of the legitimate
users of a communications line (traditionally called Alice and Bob) so
that no third party monitoring their identification can impersonate
either of them. Moreover, the system must be designed in such a way that
after a successful mutual identification even Bob cannot later on
pretend to someone else to be Alice and vice versa.

Existing identification systems are merely computationally secure, i.e.,
they rely on limited advancement of computer power, technologies, and
mathematical algorithms in the foreseeable future. The construction of a
quantum computer can seriously menace the security of classical
identification systems. A quantum identification system was first
proposed by C.~Cr\' epeau and L.~Salvail in \cite{q_mut_ident}. Their
identification protocol is based on quantum oblivious transfer
\cite{BBCS,Cr94}. Alice and Bob mutually check their knowledge of a
common secret string without disclosing it. However, quantum oblivious
transfer has been proved insecure against the so-called collective
attacks by D. Mayers \cite{M96_imp,M96_trbl}, and H.-K.~Lo and H.F.~Chau
\cite{LoChau}. Although to perform collective attacks is not possible
with current technology, recent developments suggest that it might be
possible in the near future.

In the protocols proposed here, Alice and Bob check their common secret
string in a classical way. To prevent from a later misuse, each
identification sequence is used only once and the distribution of a new
common secret string is achieved by means of quantum key distribution
(QKD). QKD, based on the BB84 protocol \cite{BB84}, has recently been
proved secure against any collective attack allowed by quantum mechanics
\cite{Biham_et_al,MayersYao}, and thus it offers unconditional
protection even against eavesdroppers possessing unlimited computational
and technological power. QKD is capable to provide two users with a
random shared secret string, whose secrecy is guaranteed by the
fundamental laws of quantum mechanics. Many papers have already been
devoted to quantum cryptography. Let us mention only a few of them
\cite{BB84,Exp92,Ekert91,BBM92,B92,ERTP92,Cr_Prg} and the survey
\cite{Br_Prg}. A large bibliography may also be found in \cite{Br_web}.

In this paper, two protocols for quantum mutual identification are presented. 
The first is designed for the case of
an unjammable
public channel. Since this requirement might appear too strong in practice,
we also
present a second protocol that utilizes unconditionally secure
authentication of messages sent over the public channel. Both protocols
have been implemented in a laboratory setup over a distance of 0.5 km.


\section{Identification with unjammable open channel}

On the assumption that the open channel used for communication during the
quantum key distribution cannot be modified, a simple identification
protocol can be implemented.
The proposed identification protocol does not rely on quantum bit
commitment or oblivious transfer, but it is based on a simple classical
identification
method using each time a new identification sequence (i.e., the sequence
is changed after each identification act, either successful or unsuccessful).
This method is secure in the following sense: a sufficient length of an
identification sequence (IS) exists such that the probability of a
success of an unauthorized user is smaller than an arbitrary small
positive number. Since each IS may be used only once, the users need
to regularly refuel their pools of sequences. Here we are coming to the
``quantum part'' of the protocol. The well known quantum key
distribution procedure (QKD) \cite{BB84} is well suitable to accomplish this
task. Of course, a certain amount of secret information must be shared at the
beginning. But later, the used ISs are replaced by new ones transmitted by
means of QKD. A limited number of ISs could be stored, e.g., on a chip card
and encrypted using a ``personal identification number'' (PIN). Owing to
discarding each used IS, the probability of a success of an unauthorized
user of the ``lost'' card depends on the number of stored ISs and on the
length of the PIN (varying these parameters, this probability could be made
arbitrarily small).

\begin{figure}
      {\begin{center}
      \epsfxsize=\hsize
      \leavevmode\epsffile{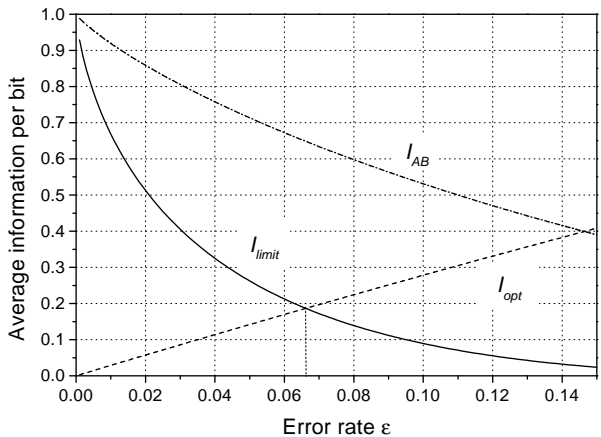}
      \end{center} }
\caption{Information as a function of error rate: $I_{AB}$ -- information
shared by Alice and Bob, $I_{opt}$ -- Eve's information gained using 
optimum eavesdropping strategy according to \protect{\cite{Fuchs}},
$I_{limit}$ -- below this limit of Eve's information on the key, the deception
probability in Protocol I can be made arbitrarily small by prolonging ISs.
The intersection of $I_{limit}$ and $I_{opt}$ shows the estimate of the
upper bound of the error rate for the identification application following
Protocol I.}
\label{graph}
\end{figure}

Let us note that there is no need to perform error correction and
privacy amplification \cite{Exp92} after QKD. The correspondence between
two compared ISs need not then be perfect, the errors being caused either by
the
imperfections of the device or by eavesdropping. If the legitimate parties
tolerate a certain
small number of errors, then it is also necessary to suppose that an
eavesdropper (Eve)
could capture some information on new ``refuelled'' ISs by measurements on
the quantum
channel. The authorized users are able to estimate the amount of this information
\cite{Fuchs}.
Nevertheless, if the ISs are long enough, this tap information is not
sufficient for Eve to succeed in the identification procedure, at least if she can
perform only separate and independent measurements on transmitted qubits
(for
the so-called coherent attack the situation is more complex).
Strictly speaking, for error rates below a certain level, the deception probability
can be made arbitrarily small by prolonging ISs. For details see
Appendix \ref{prob}.

As an example, assume an error rate $\varepsilon=0.01$ (it can be seen in
Fig.~\ref{graph} that this error rate lies below the upper bound value
$\varepsilon_{\rm
ub}$ (Eq.~\ref{upperb})). For this
error rate, the average probability that Eve correctly guesses a bit, if she applies an
optimum strategy, is approximately $\bar{p} \doteq 0.6$ (this is
computed from $I_{opt}$ -- see Fig.~\ref{graph} -- using Eq.~(65) in
\cite{Fuchs} and the definition of information). Then it follows from
Eq.~\ref{Pineq} that for ISs of length $N \ge 50$ bits, the deception
probability $P(N,\varepsilon) < 10^{-10}$.

The protocol consists of a three-pass exchange of ISs and it can be realized as
follows. Note that Alice and Bob must initially share several triads of ISs.

\begin{center}
\framebox{Protocol I (unjammable open channel)}
\end{center}

\begin{itemize}
\item Alice and Bob say each other their ordinal numbers of IS triads in the
stack  -- a pointer to the first Alice's (Bob's) unused sequence -- and choose
the higher one if they differ.
\item \begin{itemize}
      \item Alice sends the first IS of the triad to Bob.
      \item Bob checks whether it agrees with his copy. If not,
            Bob aborts communication and shifts his pointer to the next triad.
            Otherwise, he sends the second IS of the triad to Alice.
      \item Alice compares whether her and Bob's second ISs agree. If not, she
            aborts communication and shifts her pointer. Otherwise, she sends
            the third IS to Bob. If Bob finds it correct, the identification is
            successfully finished.
      \end{itemize}
\item To replace the used ISs, Alice and Bob ``refuel'' new ISs by means
      of QKD and set the pointers to their initial positions.
\end{itemize}

Three passes are necessary for the following reason: An eavesdropper
(Eve)
can pretend to be Bob and get the first IS from Alice. Of course, Alice
recognizes that Eve is not Bob because Eve cannot send the correct second
IS. So
Alice aborts connection and discards this triad (i.e., shifts the pointer to
the next one). However, later on Eve could turn to Bob and impersonate
Alice. She {\em knows\/} the first IS! Bob can recognize a dishonest Eve just
only because
she does not know the third IS.

Another possibility would be to have only one IS and to send
alternately one bit from Alice to Bob and one from Bob to Alice. The
communication is aborted when an admissible number of errors is
exceeded. However, the derivation of deception probability is more
complicated in this case.

\section{Identification with authenticated public discussion}

In practice, the ``auxiliary'' information transmitted through the open
channel during QKD {\it could \/} be modified, as it is difficult
to create a physically unjammable classical channel. Therefore authentication
of the messages sent over the open channel is necessary. This procedure
requires additional ``key'' material to be stored and transmitted in a similar
way as
ISs (again, each ``key'' may be used just once).
This authentication, however, can be utilized for the identification itself.
A three-pass authenticated public discussion, performed during QKD,
can function as the three-pass exchange of ISs described in the
preceeding section.

However, there are several problems. First, it would be more difficult to
estimate Eve's chances to succeed in the identification, if a certain number
of errors were allowed in the quantum distributed key, because the
authentication tag depends not only on the ``key'' (or IS) but also on
the message itself. So, it is necessary (or, at least, simpler) to execute
error correction and
privacy amplification.

The second problem is more subtle. For quantum cryptography to provide
unconditional security, the procedure used for authentication of public
discussion {\em must\/} also be unconditionally secure, not only
computationally. Such authentication algorithms exist \cite{Stinson}. These
algorithms are based on the so-called orthogonal arrays
\cite{BJL}. It
can be shown, however, that the length (in bits, e.g.) of an ``authentication
key'' must always be greater than the length of the authenticated
message. If $m$ is the number of all possible messages, $\kappa$ the
number of keys, and $n$ the number of all possible authentication tags,
it can be proved using methods of orthogonal arrays theory that
$$
  \kappa \ge m(n-1)+1.
$$
Now it is straightforward to show that
$$
 \kappa > m,\ {\rm if}\ n \ge 2.
$$
An example of an authentication protocol is given in Appendix
\ref{auten}.

This fact represents a difficulty for QKD. The length in bits of
the messages communicated over the public channel is always greater
than the length of transmitted ``quantum'' key. For each qubit at
least one bit of information about the basis chosen by Alice and one bit
about the basis chosen by Bob must be interchanged. Only about one half of all
successfully received qubits can be used as a key (requirement of coincidence
of bases). Besides, part of the key has to be sacrificed and compared by Alice
and Bob in order to detect potential eavesdropping, which is
also done through the open channel. So there is not enough ``quantum''
key material to replace the used bits for the next
authentication even in the case one does not intend to use the transmitted
``quantum'' key (or its part) for other purposes.

The way out of this impasse is to realize that it is not necessary
to authenticate all parts of the public discussion done during QKD.

The most important and characteristic property of quantum cryptography is
that
any attempt at eavesdropping inevitably increases the number of errors in the
transmitted key.
Thus it is necessary to prevent Eve from modifying in any way the part of
public
discussion connected with the error-rate estimation. Therefore, messages
containing the sacrificed part of the ``quantum'' key (including corresponding
bases and positions of sacrificed bits) have to be authenticated.
Any modification of the rest of public communication could impair QKD, but
would not jeopardize the security of the system. This check on error rate
should be performed as the first step of the public discussion, even before the
establishment of the {\em sifted key\/} by comparison of bases! Otherwise a
malicious Eve could manipulate the non-authenticated public transmission for
her benefit. She could, e.g., exchange separate sifted keys with Alice and
Bob
and then choose only those bits where the choice of bases coincides for all
three of them, thus obtaining full knowledge of the key without increasing
the error rate (at the cost of decreasing the transmission rate).

An important question is the length of the sacrificed subset that serves the
error-rate estimation. Alice and Bob agree on a maximum tolerable error
rate $\varepsilon_{\rm max}$, whose value must be lower than the theoretical
limit for a safe noisy quantum channel. Several such limits have been derived
in
the literature for different kinds of Eve's attacks \cite{Fuchs,Lutkenhaus},
nevertheless the ultimate value for the most general attack is not known at
present. In Appendix \ref{estim} we give a derivation of the length of the
subset and the limiting error-rate estimate $\varepsilon_{\rm lim}$ Alice and
Bob can accept so that the probability that the actual error rate is higher than
$\varepsilon_{\rm max}$, is lower than a prescribed ``safety parameter''
$\delta$.

Provided that Alice and Bob initially share a pool of secret information, the
identification procedure consists of the following steps:
\begin{center}
\framebox{Protocol II (authenticated open channel)}
\end{center}

\begin{itemize}
\item Alice and Bob first perform transmission over the quantum channel
according to the BB84 protocol, i.e., Alice randomly alternates two bases and
two bit values, while Bob records detections in randomly chosen bases (raw
quantum transmission).
\item Alice and Bob say each other their addresses in the pool of shared
secret information -- a pointer to the first Alice's (Bob's) unused bit -- and
choose the higher one if they differ. Then follows a three-pass authenticated public
discussion that serves the estimation of the error rate and mutual
identification:
        \begin{itemize}
        \item Bob sends to Alice an authenticated message containing the
                positions of bits randomly selected for error-rate estimation.
        \item  Alice checks authentication and aborts communication if it
                fails. Otherwise she sends back to Bob an authenticated
                message containing the bases and bit values of the selected
                qubits.
        	\item  Bob checks authentication and aborts communication if it fails. 
		Next he compares bases of the selected subset and retains only 
		those qubits where his and Alice's bases coincide. At last, he 
		estimates error rate. He sends to Alice	an authenticated message 
		to inform her that identification was successful and to convey 
		the value of the error-rate estimate. 
		Alice checks authentication and aborts communication if it 
		fails.
        \end{itemize}
\item If the error-rate estimate is lower than a maximum tolerable error rate $\varepsilon_{\rm lim}$, Alice and Bob compare bases of the rest of their raw data and arrive at their sifted keys. Otherwise they suspect Eve of listening in and cannot safely use the just accomplished quantum transmission to establish new shared secret sequences.
\item Then they perform error correction and privacy amplification
procedures
and arrive at an error-free distilled key. The level of privacy amplification is
based on $\varepsilon_{\rm max}$.
\item Alice and Bob refuel their shared secret information.
\end{itemize}

The used authentication sequences are always thrown away. The length of
the raw quantum transmission
must be selected such that the length of the newly obtained distilled key is
greater than the number of bits consumed for authentication/identification
purposes. It is convenient if it covers several unsuccessful identification
acts. We give concrete figures in Section~\ref{implement}.


\section{Description of the apparatus}

\begin{figure}
     {\begin{center}
      \epsfxsize=0.9\hsize
      \leavevmode\epsffile{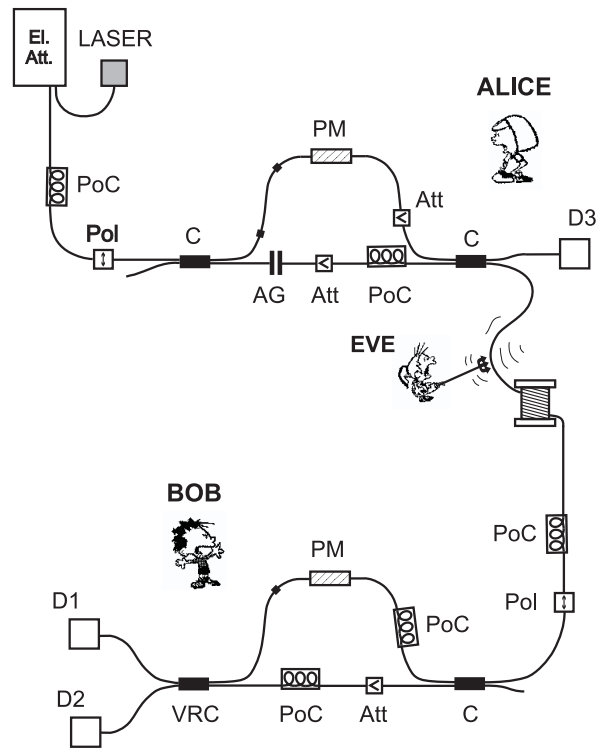}
      \end{center} }
\caption{A scheme of the optical part of the built quantum
 identification system. {\sf El.\,Att.} -- electronic attenuator,
 {\sf PoC} -- polarization controllers, {\sf PM} -- planar
 electro-optic phase modulators, {\sf ATT} -- attenuators, {\sf Pol} --
 polarizers, {\sf C} -- fibre couplers, {\sf VRC} -- variable ratio
 coupler, {\sf AG} -- air gap.}
\label{schema}
\end{figure}

The experimental implementation of our system is based on an
interferometric setup (i.e., on phase coding) with time multiplexing. It
consists of two unbalanced fibre Mach-Zehnder interferometers (see Fig.~\ref{schema}).
The path difference (2~m) of the arms of each interferometer is larger than
the
width of the laser pulse (its duration is 4~ns). Interference occurs at the
outputs of the second interferometer for pulses ``taking'' long-short
or short-long paths. These paths are of the same length and are
indistinguishable. Each of these interferometers represents the main part of
the ``terminals'' of both communicating parties. The terminals are
interconnected by a 0.5~km single mode optical fibre acting as a {\em
quantum
channel\/}, and by a classical channel (local computer network). As a light
source, a semiconductor pulsed laser with a repetition rate of 100 kHz
operating at 830~nm is used. Laser
pulses
are attenuated by a computer-controlled attenuator so that the
intensity level at the output of the first interferometer is below 1 photon
per pulse on the average. The accuracy of this setting is monitored by detector D3.
Polarization properties of light in the interferometers are controlled by
polarization controllers PoC. To balance the lengths of the arms, an air gap
AG with a remotely controlled gap-width is used. The phase coding is
performed by means of two planar electro-optic phase modulators PM (one at
each
terminal). To achieve high interference visibility, the splitting ratio of the
last combiner must approach 50:50 as closely as possible (see \cite{Budmer}).
Therefore a variable ratio coupler VRC is employed there. With this setup, it
is possible to reach visibilities well above $99.5$\%. The total losses of the
second interferometer do not exceed $4.5$~dB.

Detectors D1--D3 are single photon counting modules with Si-avalanche
photodiodes. Their output signals are processed by detection
electronics based on time-to-amplitude converters and single channel
analyzers. Both terminals are fully driven by computers. The interferometers
are
placed in polystyrene thermo-isolating boxes. Together with automatic active
stabilization of interference, it enables us to reach low error rates
(0.3--0.4~\%) with data transmission rates of the order of several kbits per
second.


\section{Practical implementation (Protocol II)} \label{implement}

To estimate the error rate of the just completed quantum transmission,
Alice and Bob sacrifice a subset of their raw data and publicly compare
bit values. It is important that this is the first step of processing the raw
data obtained from the transmission over the quantum channel. The selection of
bit positions for the subset must be completely random so that Eve has no
{\it a~priori\/} information about which bits may appear in the subset.

Let us first focus on the authenticated part of public discussion. We
choose the length of the subset $2s=2000$ bits. If the ``safety parameter'' 
$\delta=10^{-10}$ is required, we must reject all raw quantum
transmissions for which we obtain error-rate estimate $\varepsilon_{\rm est}
\ge 2.4$\% (see Appendix~\ref{estim} for details). If the total of $N$ laser
pulses have been used for the raw quantum transmission, we need
\begin{itemize}
\item $2s[\log_2 N]+a$ bits to convey and authenticate positions of selected
bits,
\item $4s+a$ bits to convey authenticated bases and bit values of the selected
bits, and,
\item say, $32+a$ bits to convey the final message whether everything is OK
or not.
\end{itemize}
Here $[x]$ denotes the smallest integer larger than
$x$, and $a\ge[\log_2 (1/\delta)]$ is the length of the authentication tag (see
Appendix~\ref{auten}). We use $a=61$.
In total this gives the requirement to share at least
\begin{equation}
b_{\rm min}=2s([\log_2 N]+2)+32+3a
\label{bmin}
\end{equation}
secret bits initially.

The length of the sifted key we obtain depends on the intensity of laser
pulses $\mu$ at the output of Alice's interferometer, on the transmissivity
of the communications line $\eta_{\rm TL}$ ($0.63$ in our device),
transmissivity of Bob's interferometer $\eta_{\rm BOB}$ ($0.35$), and the
quantum efficiency of detectors $\eta_{\rm DET}$ ($0.55$). This yields
overall transmissivity $\eta=\eta_{\rm TL} \eta_{\rm BOB} \eta_{\rm DET} =
0.12$ and we obtain a sifted key of the average length
\begin{equation}
$$N_S=\frac{1}{2} \eta \mu N.
\label{ns}
\end{equation}
The error correction and privacy amplification procedures we use are
basically those used by Bennett {\it et al.\/} \cite{Exp92}.\footnote{%
 The improved techniques of \cite{Bras1,Bras2}, enabling a more rigorous
 determination of the fraction of key bits to be discarded \cite{Lut1},
 can also be used, and very recent results of \cite{Lut2} show how to
 incorporate more sophisticated quantum non-demolition measurement
 instead of a beamsplitting attack.
}. We have
empirically found that after error correction we are left with approximately
$$
N_C = (1-2.7 \varepsilon^{2/3}) N_S
$$
bits, $\varepsilon$ being the actual error rate. At last, privacy amplification
leaves us with
 \begin{eqnarray} \nonumber
 N_D &=& N_C-\frac{\eta \mu^2}{8 \eta_{\rm TL}} N
 -\frac{2 \varepsilon_{\rm max} N_S}{\ln 2} \\ \nonumber
 &-& 5 \sqrt{N \frac{\eta \mu^2}{8 \eta_{\rm TL}} \left( 1- \frac{\eta \mu^2}{8
 \eta_{\rm TL}} \right) + \frac{2(\ln 2 +1) N_S \varepsilon_{\rm
 max}}{\ln^2 2}  }\\
 &+& \frac{\ln (\delta \ln 2)}{\ln 2}
 \label{nd}
 \end{eqnarray}
bits of distilled key\footnote{Eq.~(\ref{nd}) is valid for $\mu\ll
1$, a general relation is somewhat clumsy and we do not give it here.}. The
second term on the r.h.s.\ of Eq.~(\ref{nd}) expresses the number of bits Eve
could obtain by beamsplitting \cite{Exp92} with the capability of
replacement of
the lossy communications line by a line of $\eta_{\rm TL}=1$, while Alice
and Bob tolerate a drop in the data rate to one half of the expected
value\footnote{The intensity at Bob's side also fluctuates for several physical
reasons, therefore it is not reasonable to limit the intensity drop caused by
eavesdropping in a too restrictive way.}. The third term contains the
number of bits Eve could obtain by a probe interaction attack with the
possibility of delayed (after the announcement of the bases but before error
correction and privacy amplification) measurements on individual photons
\cite{Fuchs}\footnote{In fact, this number represents the information Eve
could obtain, which may not necessarily be in the form of a set of
deterministic bits.}. The fourth term is a 5-standard-deviations safeguard, whose 
derivation is analogous to that in \cite{Exp92}). The last term is a privacy
amplification compression that decreases Eve's information to $\delta$ bits.
Collective attacks are not included, as no bound on the information an
eavesdropper can get through a collective attack has been derived yet; it has
just been proved that such a bound exists \cite{Biham_et_al,MayersYao}.
Preliminary results on coherent eavesdropping also suggest that it does not
seem to substantially increase Eve's information \cite{Cirac}.

\begin{figure}
     {\begin{center}
      \epsfxsize=\hsize
      \leavevmode\epsffile{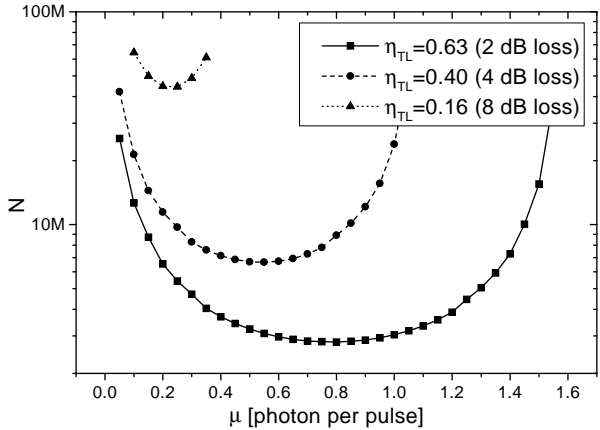}
      \end{center} }
\caption{The dependence of the number of laser pulses $N$ needed to
generate as much distilled key (Eq.~\protect{\ref{nd}}) as it is consumed for
authentication during identification (Eq.~\protect{\ref{bmin}}) on the
intensity $\mu$ of laser pulses at the output of Alice's interferometer for
three different values of the transmissivity of the communications line
$\eta_{\rm TL}$. The higher the losses of the transmission line (or its
length), the lower must be the intensity at the output
of Alice's interferometer and the greater is the number of laser pulses needed
to generate enough distilled key. We can see that the optimum
intensity is about 0.8 photon per pulse in our case ($\eta_{\rm TL}=0.63$).}
\label{r1}
\end{figure}

We have optimized this relation to maximize the ratio $N_D/N$. For our
system ($\eta=0.12$, $\varepsilon=0.004$) with the choice $\varepsilon_{\rm
max}=0.07$ and $2s=2000$), we have found an optimum average intensity $\mu
\approx 0.8$ photon per pulse (Fig.~\ref{r1}). This value represents a trade-off between
the number of pulses successfully detected by Bob and the reduction of
the length of the key caused by privacy
amplification, and sensitively depends on the overall losses of the system.
The ratio $N_D/N$ depends only weakly on $\delta$ so that it is easy to
achieve an arbitrary security level.

It is worth noting that for sufficiently low $\mu$, the
ratio $b_{\rm min}/N_D$
converges to zero with increasing $N$ so that it is always possible to
generate more
new shared secret bits than it is consumed for authentication. Therefore
authenticated QKD may be regarded as an ``expander'' of shared secret
information, once the ratio $r=N_D/b_{\rm min}$ is greater than 1. For our
system, we get $r=1$ for $N=3.1 \times 10^6$ laser pulses (see Fig.~\ref{r1}).

The whole identification procedure starts with raw quantum transmission.
In our experimental setup, we generate raw key data at sequences of
$320,000$ laser pulses. After each sequence, active stabilization of
the interferometers is performed to ensure low error rate despite
environmental perturbations. This yields an average raw key data rate of cca
5.7 kbits per second. Once about 600,000 photons are successfully detected
by
Bob (we want $r\geq2$), a three-pass authenticated public discussion is
performed according to Protocol II described above. If all three
authentications are found correct, Alice and Bob have mutually identified
themselves. If, in addition, the computed error-rate estimate falls below the
value $\varepsilon_{\rm lim}$, they are able to refuel new secret key
material.

They start doing this by comparing the bases of the rest of the raw key data,
thus arriving at approximately 300 kbits of sifted key. As final steps, they
perform error correction and privacy amplification procedures. The level of
privacy amplification is based on $\varepsilon_{\rm max}$ and the ``security
parameter'' $\delta$, as follows from Eq.~(\ref{nd}). For our usual error rates
of 0.3--0.4\%, Alice and Bob obtain about 117 kbits of distilled key generated
at an average rate of 650 bits per second. This well covers approximately
50 kbits of previously shared secret key material consumed during the
authenticated discussion. Let us note that we did not perform any special
optimization of data rate, the bottlenecks being here the way we drive the
equipment from PCs and the bandwidth of the detection electronics we used.
Nevertheless, in our setup the whole identification procedure takes
less than 3 minutes (including all auxiliary processes).


\section{Conclusions}

We have discussed the possibility to utilize the advantages of quantum
cryptography for mutual identification. Quantum key distribution can serve
as a source to ``continuously'' supply shared secret key material for
classical identification methods, which employ one key just for one
identification act. For the case of an unjammable open channel and a noisy
quantum channel, a simple identification protocol has been proposed and
deception probability has been derived. For a more realistic situation of a
jammable open channel, an identification protocol employing authentication
of
public discussion has been devised. A laboratory prototype of the
identification system has been built.
It is based on a ``one-photon'' interferometric method and on the quantum
transmission protocol BB84. The main physical parameters are following:
distance 0.5~km, wavelength 830~nm, raw data rate about 5~kbits per sec.,
distilled key generation rate 650 bits per sec., error rate in the range
0.3--0.4~\%.


\section*{Acknowledgements}

\begin{sloppypar}
The authors wish to express their gratitude to Professor Jan Pe\v rina
for his help throughout and for encouragement.
This research was supported by Czech Ministry of Education (VS 96028),
Czech Grant Agency (202/95/0002) and Czech Home Department
(19951997007,
19982003012).
\end{sloppypar}

\appendix

\section{Deception probability} \label{prob}

Let us denote $\varepsilon$ the error rate of the device. Let the length of
IS be $N$ and let us tolerate maximally $k=[\varepsilon N]$ errors in
the identification procedure ($[x]$ denotes the smallest integer greater than
$x$). If Eve's measurements are independent and if the probability that Eve
correctly guesses the $i$-th bit in the sequence is $p_i$, then the probability
that Eve succeeds in the identification is
\begin{equation}
  \label{Pdef}
  P(N,\varepsilon)= \sum_{\ell=0}^k \sum_{\{ i_1 \dots i_\ell \}}
                    \left( \prod_{j=1}^N p_j \right)
                    \left( \prod_{m=1}^\ell {q_{i_m} \over p_{i_m}}
                    \right).
\end{equation}
Here $q_i=(1-p_i)$ and the second sum goes over all
$\ell$-tuples of numbers from 1 to $N$ (for $\ell=0$ there is only $\prod_j
p_j$).
Employing Jensen's inequality \cite{Jensen}, one can find that
$$
  \prod_{j=1}^N p_j \le (\bar{p})^N,
$$
with
$$
 \bar{p} = \frac{1}{N} \sum_{j=1}^N p_j.
$$
Further, realizing that for $p \ge 1/2$ the expression $q/p \le 1$ is valid and
that for $\ell \le k$ the inequality
${N \choose \ell} \le {k \choose \ell} {N \choose k}$ holds, 
one finally obtains the deception probability:
\begin{equation}
  \label{Pineq}
  P(N,\varepsilon) \le (\bar{p})^N 2^k {N \choose k}.
\end{equation}
The question is for which $\bar{p}$ and $\varepsilon$ the
$\lim_{N\to\infty}P(N,\varepsilon)=0$ [i.e., when $P$ can be made
arbitrarily small by increasing $N$].
It can be shown that if $0<\lim_{N\to\infty}              
 \beta(N)<1$ then $\lim_{N\to\infty}               
 [\beta(N)]^N=0$ and if $\lim_{N\to\infty} \beta(N)>1$ then
$\lim_{N\to\infty}[\beta(N)]^N=\infty$
($\beta(N)$ is an arbitrary function of $N$). Thus
for each $\varepsilon$, a probability $p_{crit}$ 
\begin{equation}
 \label{pcrit}
 p_{crit}=\lim_{n\to\infty} 2^{-k/n} {n \choose k}^{-1/n},
\end{equation}
may be defined
such that for all $\bar{p}<p_{crit}$, the limit
$\lim_{N\to\infty}P(N,\varepsilon)=0$.
The graph in Fig.~\ref{graph} shows average information per
bit $I_{limit}=1 + p_{crit} \log_2(p_{crit})  + (1-p_{crit})
\log_2(1-p_{crit}) $ corresponding to $p_{crit}$ together with mutual
information of Alice and Bob $I_{AB}$ and Eve's information gained by
optimal
eavesdropping strategy $I_{opt}=I_{AE}=I_{EB}$ (see \cite{Fuchs}
Eq.~(65)) as a function of error rate $\varepsilon$. The intersection
of $I_{limit}$ and $I_{opt}$ determines the estimation of the upper bound of
the error rate for this identification application:
\begin{equation}
 \label{upperb}
 \varepsilon_{ub} \approx 0.066.
\end{equation}

A disputative question might be the case of collective (or coherent)
attacks when the probabilities of the correct guesses of particular bits need not be
independent anymore and then the probability $P(N,\varepsilon)$ may
decrease with increasing $N$ more slowly in comparison with the previous
case.

\section{Example of authentication protocol} \label{auten}

If probabilities of impersonification are to be the same for all
possible pairs {\it (message, authentication tag)}, then there exists an
orthogonal array that serves as a base for an authentication code. In such a
case the deception probability, defined as a maximum
from the above mentioned probabilities of impersonification, is minimal and
is equal to the reciprocal of the number of all possible
authentication tags.

There is a class of orthogonal arrays that enables us to construct
reasonable authentication codes \cite{Stinson}. If $p$ is prime and $d \ge
2$ is an integer, an authentication code can be created for $(p^d -
1)/(p-1)$ messages with $p^d$ keys and $p$ authentication tags
(the deception probability is $p^{-1}$). For a given message and
a given authentication key, the authentication tag can be calculated as
follows:
\begin{enumerate}
 \item
   Convert a given authentication key to the number system of the
   base $p$ (its maximal length in this system is $d$). Let us denote
   the $i$-th ``digit'' by $r_i$.
 \item
   Construct and order all non-zero ``numbers'' in the number system of
   the base $p$ of the maximum length $d$ that have the first non-zero
   ``digit'' from the left equal to 1 [there are $(p^d - 1)/(p-1)$ such
   numbers]. A one-to-one mapping exists between all possible messages
   and all ``numbers'' (or sequences) from this set. Assign the
   corresponding ``number'' (the ordering of the ``numbers'' is assumed to
   be fixed) to the message to be authenticated. Let the $i$-th
   ``digit'' of that particular ``number'' be denoted by $c_i$.
 \item
   The authentication tag is given by the equation
   $$
     A(r,c) = \sum_{i=1}^{d} r_i c_i \ {\rm mod}\ p.
   $$
\end{enumerate}

As a practical example (used in implementation of Protocol II), we have
chosen $p=2^{61}-1$ (it's prime) and $d=739$. Then the deception probability
is about $5 \cdot 10^{-19}$. The length of the key is 45079 bits, the length of
the message can be up to 45017 bits and the authentication tag consist of 61
bits.

By the way, in case of $p$ of this form it is not necessary to make
the conversion of item (1) above. One can just create groups consisting
of 61 random bits. What is only necessary is to discard groups
containing all 61 ones (the probability of appearance of such a group is
deuced small).

\section{Derivation of the length of subset for error-rate estimation}
\label{estim}

\begin{figure}[b]
     {\begin{center}
      \epsfxsize=\hsize
      \leavevmode\epsffile{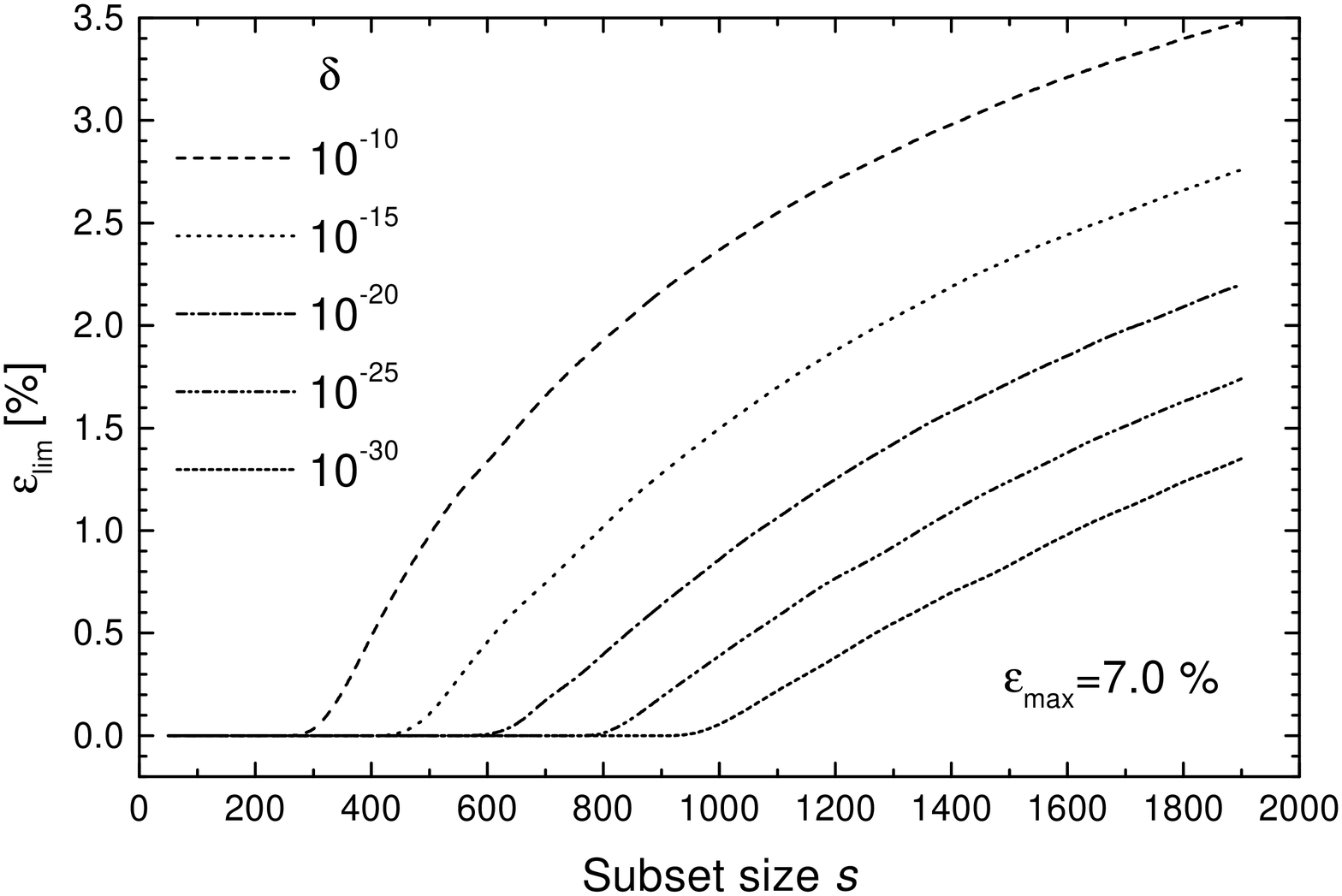}
      \end{center} }
\caption{The dependence of limiting value $\varepsilon_{\rm lim}$ on a
subset size $s$ for different values of the ``security parameter'' $\delta$,
when a maximum error rate of $\varepsilon_{\rm max}=0.07$ is tolerated (see
Eq.~\protect{\ref{emax}}). A subset of the length $2s$ is randomly selected
from raw quantum data which yields $s$ bits with coincident bases on the
average. Quantum transmission is considered insecure (i.e., the probability of
the actual error rate $\varepsilon$ being higher than $\varepsilon_{\rm max}$
is not lower than $\delta$), if the error-rate estimate $\varepsilon_{\rm est}$
obtained from the subset check exceeds the value $\varepsilon_{\rm lim}$. In
our case we choose $s=1000$ and $\delta=10^{-10}$, and we find
$\varepsilon_{\rm lim} \approx 2.4$\%.}
\label{limit}
\end{figure}

Let us suppose that we select a subset of length $2s$. After
comparison of bases, $s$ bits will be retained on the average.
Provided that the actual error rate is $\varepsilon$, the probability that
we find $k$ errors in the subset of length $s$ (i.e., the error-rate
estimate is $\varepsilon_{\rm est}=k/s$) is given by
\begin{equation}
p(\varepsilon_{\rm est}|\varepsilon)={s \choose k} \varepsilon^k
(1-\varepsilon)^{s-k}.
\end{equation}
Applying Bayes' theorem, the probability that the actual error rate is
$\varepsilon$, when the estimate is $\varepsilon_{\rm est}=k/s$, is given by
\begin{equation}
p(\varepsilon|\varepsilon_{\rm est})={{\left\{ \varepsilon^{\varepsilon_{\rm
est}} (1-\varepsilon)^{1-\varepsilon_{\rm est}} \right\}^s} \over
{\int_{0}^{1}  \left\{ \varepsilon^{\varepsilon_{\rm est}}
(1-\varepsilon)^{1-\varepsilon_{\rm est}} \right\}^s} d\varepsilon}.
\end{equation}
Here we assume a uniform distribution of $\varepsilon$.
We are now interested in finding a
limiting value $\varepsilon_{\rm lim}$ such that for all $\varepsilon_{\rm est}
\le \varepsilon_{\rm lim}$ the probability
\begin{equation}
P(\varepsilon>\varepsilon_{\rm max}) =
\int_{\varepsilon_{\rm max}}^{1} p(\varepsilon|\varepsilon_{\rm est})
d\varepsilon \le \delta,
\end{equation}
where a small positive number $\delta$ denotes the ``security parameter''. In
Fig.~\ref{limit} we plot the solution of the equation
\begin{equation}
{\int_{\varepsilon_{\rm max}}^{1} {\left\{ \varepsilon^{\varepsilon_{\rm
lim}} (1-\varepsilon)^{1-\varepsilon_{\rm lim}} \right\}^s} d\varepsilon \over
{\int_{0}^{1}  \left\{ \varepsilon^{\varepsilon_{\rm lim}}
(1-\varepsilon)^{1-\varepsilon_{\rm lim}} \right\}^s} d\varepsilon} = \delta
\label{emax}
\end{equation}
with respect to $\varepsilon_{\rm lim}$ for several values of $\delta$.
A maximum acceptable error rate
$\varepsilon_{\rm max}=0.07$ has been chosen, which is well below the
lowest
security limit derived so far (0.146) \cite{Fuchs}. The graph in
Fig.~\ref{limit} should be understood as follows: Once we select a suitable
value for the subset length, $s$, and the ``security parameter'' $\delta$, the
corresponding curve suggests a limiting value for the estimated error level,
above which the transmitted sequence should be rejected as it cannot be
guaranteed to have the actual error rate $\varepsilon \le \varepsilon_{\rm max}$
with the required probability $1-\delta$.

\begin{flushleft}

\end{flushleft}

\end{document}